\documentstyle{article}
\begin{document}
\begin{titlepage}
\large

\begin{center}
National Academy of Sciences of Ukraine\\
Institute for Condensed Matter Physics
\end{center}

\vspace{2cm}
\hspace{6cm}
\begin{tabular}{l}
Preprint \\
ICMP - 96 -  26E
\end{tabular}

\vspace{2cm}

\begin{center}
S.I.Sorokov, R.R.Levitskii, T.M.Verkholyak
\end{center}

\begin{center}
Investigation of the annealed disordered Ising systems within
two-tail approximation
\end{center}

\vspace{1.5cm}

\begin{abstract}
{In the present paper an approach for investigation of the disordered
two-component Ising systems with long range interaction has been suggested.
Possible applications to metalic and magnetic alloys and lattice gas
are considered. We have also obtained numerical results for thermodynamical
properties of these models. The comparison of numerical results obtained
within mean field, gaussian field and two-tail
approximations are carried out.}
\end{abstract}

\vspace{2cm}

\begin{center}
L'viv - 1996
\end{center}
\end{titlepage}

\section{Introduction}
The present paper is devoted to the investigation of a binary annealed
disordered magnet on the basis of Ising model. This model and its simpler
versions are often used in the study of metalic alloys \cite{yg,gk,kha},
magnetic alloys \cite{vz,kaw}, ferroelectric solutions, in particular,
the ferroelectrics with hydrogen bonds \cite{lss}.

Let us point out several works concerning the magnetic alloys. In the
work \cite{vz}, V.G.Vaks and N.E.Zein obtained phase diagrams for the model
of the binary magnetic alloy with non-magnetic impurities within two
particle cluster approximation at $T>T_m$ ($T_m$ is the temperature of the
magnetic ordering). It was noted that in mean field approximation (MFA)
in contrast to cluster
approximation the magnetic subsystem does not influence the phase diag\-ram.
Later by means of MFA T.Kawasaki \cite{kaw} investigated the influence of
the magnetic subsystem on the properties of the atomic subsystem and v.v. at
$T<T_m$. In the work \cite{lss} a detailed investigation of
this model within two-site cluster approximation were performed. There
was found the influence of the spin subsystem on binodal (coexistence
temperature $T_b$) and spinodal (spontaneous separation temperature $T_s$)
temperatures of the system in a spin ordering phase. Here the difference
between quenched and annealed types of disorder has been investigated and
correlation functions have been calculated.

The non-magnetic version of the model is often applied to the
study of real binary alloys \cite{yg,gk,kha}. Using the pseudopotential
theory for defining of intersite interaction Z.Gurskii and Yu.Khokhlov tried
to explain the properties of Ca-Ba and K-Cs alloys \cite{gk}. For
the thermo\-dynamical averaging they used the collective variables method
with a non-selfconsistent equation for unknown chemical potential.

This work has been directed to the study of a disordered model with
long range interaction or short range one in a case of large $z$ ($z$ is the
number of nearest neighbours). At first we consider the Hamiltonian of
$M$-component Ising model and study the model within mean field
approximation (MFA), Gaussian field approximation (GFA) and two-tail
approximation (TTA). But
for the sake of simplicity all approximations are formulated for the case
$M=2$.

%\setlength{\mathindent}{0cm}

%\section*{2. Description of the model }
\section{Description of the model }
\setcounter{equation}{0}
\renewcommand{\theequation}{\arabic{section}.\arabic{equation}}

Let us consider $M$-component system with site disorder on Bravais lattice
with the following Hamiltonian:
% 1
\begin{eqnarray}
&&
-\beta H = {\cal H} = {\cal H}_x + {\cal H}_s,
\end{eqnarray}
where
% 2, 3
\begin{eqnarray}
&& {\cal H}_x = \sum_{\alpha=1}^M   \sum_{i = 1}^N
\varepsilon_{\alpha} x_{i \alpha} + \frac{1}{2} \sum_{\alpha,\beta = 1}^M
\sum_{i,j=1}^N V_{\alpha\beta}(i-j) x_{i \alpha} x_{j\beta}, \\
%\end{eqnarray}
%\begin{eqnarray}
&& {\cal H}_s = \sum_{\alpha = 1}^M \sum_{i = 1}^N
\Gamma_{\alpha} x_{i\alpha} S_{i\alpha} + \frac{1}{2}
%\stackrel{M}{\sum}
\sum_{\alpha,\beta = 1}^M
%\stackrel{N}{\sum}
\sum_{i,j=1}^N {\cal I}_{\alpha\beta}(i-j) x_{i \alpha}
S_{i\alpha}x_{j \beta} S_{j\beta}.
\end{eqnarray}

All parameters of the Hamiltonian (2.1) contain the factor $(-\beta)$.
%that will be picked out in the final expressions.
The Hamiltonian ${\cal
H}_x$ corresponds to the disordered ionic subsystem. Here
$\varepsilon_{\alpha}=\vartheta_{\alpha}+\mu_{\alpha}$,
$\mu_{\alpha}=-\beta\mu^R_{\alpha}$ is the
dimensionless chemical potential of $\alpha$-type ions,
$\vartheta_{\alpha}$ is related to the difference between alloy component
ion characteristics;
$V_{\alpha\beta}(i-j)=V_{\alpha\beta}(R_{ij})$ is the potential of effective
ionic interaction; $x_{i\alpha}=1$, if the site is occupied by the ion of
type $\alpha$ and 0 otherwise. The Hamiltonian ${\cal H}_s$ corresponds to
the system of $N$ Ising spins ($S_{i\alpha}=S^z_{i\alpha}$) situated on the
ions with external field energy $h_{\alpha}=(
{\mbox{\boldmath$\mu$}}_{\alpha}
{\bf h})$ and
exchange interaction
${\cal I}_{\alpha\beta}(i-j)={\cal I}_{\alpha\beta}(R_{ij})$.

In the present paper we confine ourselves by the case of binary alloy
$(M=2)$. It should be noted that suggested approximations below can be
applied to magnetic alloys with arbitrary number of sorts. It is
convenient to use  the spin variables instead of $x_{i1}, x_{i2}
(\alpha =1,2)$
%
% 1
\begin{eqnarray}
&&
x_{i\alpha} = \frac{1-(-1)^{\alpha} S_{i0}}{2};
\quad S_{i0} = x_{i1}-x_{i2}.
\end{eqnarray}
Now the Hamiltonian (2.1) can be written as follows
%
% 2
\begin{eqnarray}
&&
{\cal H}_{xs} = E({\mbox{\boldmath$\varepsilon$}}) +
{\cal H}_{\Theta} ({\mbox{\boldmath$\Gamma$}}),
\\
&&
{\cal H}_{\Theta} ({\mbox{\boldmath$\Gamma$}}) =
\sum_{\sigma = 0,1,2}\sum_{i=1}^N
\Gamma_{\sigma} \Theta_{i,\sigma}+ \frac{1}{2}
%\sum_{\sigma, \sigma' \atop i,j}
\sum_{\sigma, \sigma'}\sum_{i,j=1}^N
{\cal I}_{\sigma, \sigma '}(R_{ij}) \Theta_{i\sigma} \Theta_{j\sigma}.
\nonumber
\end{eqnarray}
Here we use the following notations
%
% 3
\begin{eqnarray}
&&
{\mbox{\boldmath$\Gamma$}}  =
\left(
\begin{array}{c}
\Gamma_0 \\
h_1 \\
h_2
\end{array} \right),
\quad
{\Theta }_i =
\left(
\begin{array}{c}
S_{i0} \\
x_{i1}S_{i1} \\
x_{i2}S_{i2}
\end{array} \right),
\quad
\nonumber\\
&&
\hat{\cal I} (R) =
\left(
\begin{array}{ccc}
{\cal I}(R) & 0 & 0 \\
0 & {\cal I}_{11}(R) & {\cal I}_{12} (R) \\
0 & {\cal I}_{12}(R) & {\cal I}_{22} (R)
\end{array} \right),
\nonumber \\
&&
\frac{1}{N} E = \frac{1}{2} (\varepsilon_1 + \varepsilon_2) +
\frac{1}{8}(V_{11} + V_{22} + 2V_{12}),
\\
&&
\Gamma_0 = \frac{1}{2} (\varepsilon_1 - \varepsilon_2) + \frac{1}{4}
(V_{11} - V_{22}) \;
(V_{\alpha,\beta}=V_{\alpha,\beta}(q\rightarrow 0)),
\nonumber \\
&&
4{\cal I}(R) = V_{11} (R) + V_{22} (R) - 2 V_{12} (R).
\nonumber
\end{eqnarray}
The Hamiltonian ${\cal H}_{\Theta}({\mbox{\boldmath$\Gamma$}} )$ describes
the three-sort spin system. The ther\-mo\-dynamical po\-ten\-ti\-al (TP) of
system (2.1) has the following form
(${\bf h}=(h_1,h_2)$)
%
% 2.7
\begin{eqnarray}
&&
\Omega ({\mbox{\boldmath $\varepsilon$}},{\bf h}) = - \beta^{-1}
\ln Sp_{\{s,x\}}~e^{{\cal H}(x,s)} =
E(\varepsilon) + \Omega_{\Theta} ({\mbox{\boldmath$\Gamma$}}).
\end{eqnarray}
Here $\Omega_{\Theta} ({\mbox{\boldmath$\Gamma$}})$ is the TP for the
Hamiltonian ${\cal H}_{\Theta}({\mbox{\boldmath$\Gamma$}})$.
The correlation functions $(CFs)$ of an arbitrary order can be obtained on
the basis of the relations
%
% 2.8
\begin{equation}
\Omega_{\Theta}^{(l)}(i_1 \delta_1, \ldots, i_l \delta_l) = \langle
\Theta_{i_1 \delta_1}\cdots \Theta_{i_l \delta_l} {\rangle}^c =
\frac{\delta}{\delta \Gamma_{i_1 \delta_1}} \cdots  \frac{\delta}{\delta
\Gamma_{i_l \delta_l}}[- \beta \Omega_{\Theta} (\Gamma )].
\end{equation}
Here we have replaced the uniform fields $\Gamma_{\alpha}$ with nonuniform
ones $\Gamma_{i\alpha}$. The TP (2.7) is a function of temperature $(T)$
and chemical potentials $(\mu_1, \mu_2)$. Transition to concentration
variables $c_{\alpha}$ instead of $\mu_{\alpha}$ can be performed by the
Legendre transformation
%
% 2.9
\begin{eqnarray}
&&
\frac{1}{N} F ({\bf c, h})
= \frac{1}{N} \Omega ({\mbox{\boldmath $\varepsilon$}},{\bf h}) -
\sum_{\alpha=1,2} \mu^R_{\alpha} c_{\alpha}.
\end{eqnarray}
Here $F ({\bf c, h})$ is a free energy of system  (2.1)
and chemical potentials
are defined by $(m_0 = \langle S_{i0}\rangle )$
%
% 2.10
\begin{eqnarray}
c_{\alpha} = \frac{N_{\alpha}}{N} = \frac{\delta}{\delta
\varepsilon_{\alpha}} [-\beta\Omega (\varepsilon, h)] = \frac{1-
(-1)^{\alpha} m_0}{2}.
\end{eqnarray}
The system (2.10) gives only one independent equation for  $\mu_1 -\mu_2$ or
$\Gamma_0$, because $\sum_{\alpha}c_{\alpha}=1$.
%(see (2.4)).
Here we consider the system which undergoes both
the separation phase transition and magnetic phase at temperature
$T_m$. Therefore, the expression for $\langle S_{i0} \rangle$ can contain
the magnetic moments $m_1 =\langle S_{i1} \rangle$ and $m_2 =\langle
S_{i2}\rangle$ as well. The equations for them we will obtain proceeding
from (2.8):%
% 2.11
\begin{eqnarray}
&&
m_1 = \frac{\delta}{\delta \Gamma_1} \cdot \frac{1}{N} [-\beta
\Omega_{\Theta} ({\mbox{\boldmath $\Gamma$}})],
\quad
m_2 = \frac{\delta}{\delta \Gamma_2} \cdot \frac{1}{N} [-\beta
\Omega_{\Theta} ({\mbox{\boldmath $\Gamma$}})].
\end{eqnarray}
 The system of equations (2.10), (2.11) gives the values for $\Gamma_0$,
$m_1$, $m_2$ as function of tempe\-rature $T$ and concentration $c=c_1$.
The spontaneous separation transition occurs at $T_s$ (spinodal temperature)
which is found as divergence temperature of CF $(\langle
S_{i0}S_{j0}{\rangle}^c)_{{\bf q} = 0}$. The coexistence temperature $T_b$
(binodal temperature) of both phases A and B is calcu\-la\-ted from the
following system
%
% 2.12
\begin{eqnarray}
&&
\left \{
\begin{array}{l}
\Omega ({\mbox{\boldmath $\varepsilon$}}(m_0),{\bf h},T_b) /_{m_0=m_{0A}} =
\Omega ({\mbox{\boldmath $\varepsilon$}}(m_0),{\bf h},T_b) /_{m_0=m_{0B}} \\
\Gamma_0 (m_{0A},{\bf h}, T_b) = \Gamma_0 (m_{0B},{\bf h}, T_b)
\end{array}
\right.
%\nonumber
\end{eqnarray}
The system (2.12) gives the dependences $m_{0A} (T_b),m_{0B}(T_b)$ and it is
solved jointly with system (2.10), (2.11).

Although for the system (2.12) exact solution has not been found, one can
make some conclusions about the symmetry of phase diagrams. In case
${\cal I}_{11}={\cal I}_{22}$ and $\Gamma_1=\Gamma_2$ the thermodynamical
potential
\begin{eqnarray}
&&
\Omega_{\Theta} (\Gamma_0, {\mbox {\bf h}})=
-\beta^{-1}\ln {\mbox Sp}\exp{} \!\!\left[
\sum_{i=1}^N \Gamma_0(x_{i1}{-}x_{i2}){+} \right.
\nonumber\\
&&
\left.
\frac{1}{2}\sum_{i,j=1}^N{\cal I}(i{-}j)(x_{i1}{-}x_{i2})(x_{j1}{-}x_{j2}){+}
{\cal H}_s
\right]=\Omega_{\Theta} (-\Gamma_0, {\mbox {\bf h}})
\nonumber
\end{eqnarray}
will be even function of $\Gamma_0$ (because trace over $x_{i1}$ and
$x_{i2}$ becomes equivalent). Similarly, it can be proved that
\begin{eqnarray}
&&
m_0(\Gamma_0)=\langle x_{i1}{-}x_{i2}\rangle=
\nonumber\\
&&
\frac
{{\mbox Sp}(x_{i1}{-}x_{i2})
{\mbox e}^{
{\sum} \Gamma_0(x_{i1}{-}x_{i2}){+}
\frac{1}{2}{\sum}{\cal I}(i{-}j)(x_{i1}{-}x_{i2})(x_{j1}{-}x_{j2}){+}
{\cal H}_s
}
}
{{\mbox Sp}
{\mbox e}^{
\sum\Gamma_0(x_{i1}{-}x_{i2}){+}
\frac{1}{2}{\cal I}(i{-}j)(x_{i1}{-}x_{i2})(x_{j1}{-}x_{j2}){+}
{\cal H}_s
}
}=
\nonumber\\
&&
\frac
{{\mbox Sp}(x_{i2}{-}x_{i1})
{\mbox e}^{
{\sum} (-\Gamma_0)(x_{i1}{-}x_{i2}){+}
\frac{1}{2}{\sum}{\cal I}(i{-}j)(x_{i1}{-}x_{i2})(x_{j1}{-}x_{j2}){+}
{\cal H}_s
}
}
{{\mbox Sp}
{\mbox e}^{
\sum\Gamma_0(x_{i1}{-}x_{i2}){+}
\frac{1}{2}{\cal I}(i{-}j)(x_{i1}{-}x_{i2})(x_{j1}{-}x_{j2}){+}
{\cal H}_s
}
}
\nonumber\\
&&
=\langle x_{i2}{-}x_{i1}\rangle=-m_0(-\Gamma_0)
\nonumber
\end{eqnarray}
is odd function of $\Gamma_0$, where $\Gamma_0\sim (\mu_1-\mu_2)$
and determined by the formula (2.6). This means that $\Omega_{\Theta}$ is
even function relatively variable $m_0$ and first equation of the system
(2.12) has following solution $m_{0B}=-m_{0A}$. Then second equation for
$m_{0}$ can be rewritten as following:
$\Gamma_0 (m_{0A},{\bf h}, T_b) = 0$.
Therefore, phase diagram is symmetrical relatively variables $m_0$ or
permutation of components' concentration $c_1$ and $c_2$.

The spinodal temperature also is symmetrical function of $m_0=c_1-c_2$
in case ${\cal I}_{11}={\cal I}_{22}$ and
$\Gamma_1=\Gamma_2$. Spinodal decay takes place if the CF
$(\langle
S_{i0}S_{j0}{\rangle}^c)_{{\bf q} = 0}$ diverges. It means that
the system becomes unstable against the infinitesimal
fluctuation of concentration: $\frac{\partial^2{F}}{\partial m_0^2} =
0$. Since $\frac{\partial{F}}{\partial m_0} = \Gamma_0$ , the last
condition may be rewritten as $\frac{\partial \Gamma_0}{\partial m_0} = 0$.
$\Gamma_0$ is odd function of $m_0$, then
$\frac{\partial \Gamma_0}{\partial m_0}$ will be even function of $m_0$.
This means if for a certain temperature $T_s$ the equation
$\frac{\partial \Gamma_0}{\partial m_0}=0$
has solution $m_{0}$, the solution $-m_{0}$ also exists and spinodal
temperature $T_s$ is an even function relatively $m_0$.

Further for calculation of TP and
CFs of system with Hamiltonian ${\cal H}_{\Theta} ({\mbox{\boldmath
$\Gamma$}})$ we will use the expansion over  $1/z$ and the results of ref.
\cite{ylsd}. Now the TP $\Omega_{\Theta} ({\mbox{\boldmath $\Gamma$}})$ can
be written as follows
$({\cal I}_{\alpha \beta} =({\cal I}_{\alpha \beta}({\bf
q} \to 0),V= V ({\bf q} \to 0))$
%
% 2.13, 2.14
\begin{eqnarray}
&& - \frac{\beta}{N} \Omega_{\Theta} ({\mbox{\boldmath $\Gamma$}}) =
- \frac{1}{2} \sum_{\delta,\delta{'}=0,1,2} {\cal I}_{\delta\delta{'}}
m_{\delta} m_{\delta{'}} + f({\mbox{\boldmath $\ae$}}), \\
&& f({\mbox{\boldmath $\ae$}}) = \langle F^{(0)} ({\mbox{\boldmath $\ae$}} +
{\mbox{\boldmath $\sigma$}}) \rangle -\frac{1}{2} \sum_{\delta,\delta{'}
=0,1,2}
\lambda^{(2)}_{\delta \delta{'}} \langle F^{(2)}_{\delta \delta{'}}
({\mbox{\boldmath $\ae$}} +{\mbox{\boldmath $\sigma$}}) \rangle -
\nonumber\\
&& -\frac{1}{2N} \sum_{q} \ln \det [1 -
\hat {\cal I} (q) \langle \hat F^{(2)}_{\delta \delta{'}} ({\mbox{\boldmath
$\ae$}} +{\mbox{\boldmath $\sigma$}}) \rangle ].
%\nonumber
\end{eqnarray}
Here irreducible part $f({\mbox{\boldmath $\ae$}})$ of $\Omega_{\Theta}
(\Gamma)$ is written with the accuracy up one-loop diagrams (two-tail
diagrams approximation (TTA)).

We use the following notations from ref. \cite{ylsd} for averages
of arbitrary function
$y ({\mbox{\boldmath $\ae$}} + {\mbox{\boldmath $\sigma$}})$  over
fluctuating fields {\mbox{\boldmath $\sigma$}} with Gaussian distribution
function $\rho_2({\mbox{\boldmath $\sigma$}})$:
%
% 2.15
\begin{eqnarray}
&&
\langle y ({\mbox{\boldmath $\ae$}} + {\mbox{\boldmath $\sigma$}})
\rangle = \int d \sigma_0 d \sigma_1d \sigma_2 \rho_2 ({\mbox{\boldmath
$\sigma$}}) y ({\mbox{\boldmath $\ae$}} + {\mbox{\boldmath $\sigma$}}),\\
&&
\rho_2 ({\mbox{\boldmath $\sigma$}}) = [\det 2 \pi \hat \lambda^{(2)}]^{-1/2}
\exp \{ - \frac{1}{2} \sum_{\delta,\delta{'}}
[\hat \lambda^{(2)}]^{-1}_{\delta\delta{'}}
\sigma_{\delta} \sigma_{\delta{'}} \},
\nonumber \\
&&
\ae_0 = \Gamma_0 + {\cal I} m_0, \quad \ae_{\alpha} = h_{\alpha}
+ \sum_{\beta}
{\cal I}_{\alpha\beta} m_{\beta}~(\alpha,\beta =1,2),
\nonumber
\end{eqnarray}
where $[\hat \lambda^{(2)}]^{-1}$ is the inverse matrix to
$\hat \lambda^{(2)}$.
In (2.15) we use the following notations for functions on
${\bf x} =(x_0, x_1, x_2)$
%
% 2.16
\begin{eqnarray}
&& F^{(0)} ({\bf x}) = \ln Z ({\bf x}) = \ln \{ e^{x_0} Z_1 (x_1) + e^{-
x_0}Z_2 (x_2) \};
\nonumber \\
&&  F^{(1)}_{\delta} ({\bf x}) = \frac{\delta}{\delta x_{\delta}} F^{(0)}
({\bf x});
~F^{(2)}_{\delta\delta{'}} ({\bf x}) = \frac{\delta}{\delta x_{\delta '}}
\cdot \frac{\delta}{\delta x_{\delta}} F^{(0)} ({\bf x}); \\
&& Z_{\alpha} (x_{\alpha}) = Sp_{s_{\alpha}} e^{x_{\alpha} s_{\alpha}}
\mathop{\Longrightarrow}\limits_{s_{\alpha} = \pm 1}\relax
  2 \cosh x_{\alpha};
\nonumber\\
&& {\cal F}^{(0)}_{\alpha}(x_{\alpha})=\ln Z_{\alpha} (x_{\alpha});
~{\cal F}^{(l)}_{\alpha}(x_{\alpha})=\frac{\delta^l}{\delta x^l_{\alpha}}
{\cal F}^{(0)}_{\alpha}(x_{\alpha}).
\nonumber
\end{eqnarray}

From stationarity conditions with respect to nine  variables $m_0,
m_{\alpha}$,
$\lambda^{(2)}_{00}$, $\lambda^{(2)}_{0\alpha} = \lambda^{(2)}_{\alpha 0}$,
$\lambda^{(2)}_{\alpha\beta}$ we find the system of nine equations for nine
unknown $\Gamma_0, m_{\alpha}$,
$\lambda^{(2)}_{00}$, $\lambda^{(2)}_{0\alpha} = \lambda^{(2)}_{\alpha 0}$,
$\lambda^{(2)}_{\alpha\beta}$
%
% 2.17, 2.18
\begin{eqnarray}
&& 2c-1 =m_{0} = \langle F^{(1)}_0 ({\mbox{\boldmath $\ae$}} +
{\mbox{\boldmath $\sigma$}}) \rangle,~m_{\alpha} = \langle F^{(1)}_{\alpha}
({\mbox{\boldmath $\ae$}} + {\mbox{\boldmath $\sigma$}}) \rangle, \\
&& \lambda^{(2)}_{\delta\delta{'}} =
\left( \frac{1}{2N} \sum_{q} [\hat 1 -
\hat{\cal I} (q)\langle F^{(2)} ({\mbox{\boldmath $\ae$}} + {\mbox{\boldmath
$\sigma$}}) \rangle ]^{-1} \hat{\cal I} (q) \right)_{\delta\delta{'}}.
\end{eqnarray}
In the works of Onyszkiewicz at all \cite{onysz}
GFA was suggested. It
can be obtained expanding
$\ln \det [\hat 1 -
\hat {\cal I} (q) \langle \hat F^{(2)}_{\delta \delta{'}} (
{\mbox{\boldmath $\ae$}}+
{\mbox{\boldmath $\sigma$}}) \rangle ]
$
in (2.14)
and
$\left( \frac{1}{2N} \sum_{q} [\hat 1 - \right.$ $\left.
\hat{\cal I} (q)\langle \hat F^{(2)} ({\mbox{\boldmath $\ae$}} +
{\mbox{\boldmath
$\sigma$}}) \rangle ]^{-1} \hat{\cal I} (q) \right)_{\delta\delta{'}}
$
in (2.18) in ${\cal I}(q)$ up to the terms of second
order. In the result :
%2.19
\begin{eqnarray}
f({\mbox{\boldmath $\ae$}}) = \langle F^{(0)} ({\mbox{\boldmath $\ae$}} +
{\mbox{\boldmath $\sigma$}}) \rangle -\frac{1}{2}
\sum_{\delta,\delta{'} =0,1,2}
\lambda^{(2)}_{\delta \delta{'}} \langle F^{(2)}_{\delta \delta{'}}
({\mbox{\boldmath $\ae$}} +{\mbox{\boldmath $\sigma$}}) \rangle -
\nonumber\\
-\frac{1}{2N}
\sum_{\delta,\delta_1,\delta_2=0,1,2}
\langle F^{(2)}_{\delta_1 \delta{'}} ({\mbox{\boldmath
$\ae$}} +{\mbox{\boldmath $\sigma$}}) \rangle
\langle F^{(2)}_{\delta_2 \delta} ({\mbox{\boldmath
$\ae$}} +{\mbox{\boldmath $\sigma$}}) \rangle
\sum_{q} {\cal I}_{\delta \delta_1} (q) {\cal I}_{\delta{'} \delta_2} (q),\\
%\nonumber
\lambda^{(2)}_{\delta\delta{'}} =
\frac{1}{2N} \sum_{\delta_1,\delta_2=0,1,2}
\langle F^{(2)}_{\delta_1 \delta_2} ({\mbox{\boldmath
$\ae$}} +{\mbox{\boldmath $\sigma$}}) \rangle
\sum_{q} {\cal I}_{\delta \delta_1} (q) {\cal I}_{\delta_2 \delta{'}} (q).
\end{eqnarray}

%%%Molecular field approximation
In contrast to GFA and TTA MFA permits to get analytical results for
thermodynamical functions and transition temperatures in case
of non-magnetic alloys. This approximation one can obtain neglecting
fluctuation of molecular field $m_{\delta}$. The expression for TP is
% 2..., 2....
\begin{eqnarray}
&& - \frac{\beta}{N} \Omega_{\Theta} ({\mbox{\boldmath $\Gamma$}}) =
- \frac{1}{2} \sum_{\delta,\delta{'}=0,1,2} {\cal I}_{\delta\delta{'}}
m_{\delta} m_{\delta{'}} + F^{(0)}({\mbox{\boldmath $\ae$}}).
\end{eqnarray}
The equation for chemical potential and order parameters is determined
by the relations (2.10), (2.11). In MFA one can obtain:
\begin{eqnarray}
&&
2c_1{-}1{=}m_0{=}F_0^{(1)}({\mbox{\boldmath $\ae$}}){=}
\left[
{\mbox e}^{\ae_0}Z_1(\ae_1){-}{\mbox e}^{-\ae_0}Z_2(\ae_2)
\right]
{Z({\mbox{\boldmath $\ae$}})}^{-1},\\
%\nonumber\\
&&
m_{\alpha}=F_{\alpha}^{(1)}({\mbox{\boldmath $\ae$}})=
{\mbox e}^{(-1)^{\alpha+1}\ae_0}Z_{\alpha}^{(1)}(\ae_{\alpha})
{Z({\mbox{\boldmath $\ae$}})}^{-1}.
\end{eqnarray}
The equation for $m_0$ can be transformed to two dependent equations for
the components' concentration $c_{\alpha}$ (see 2.10)):
\begin{eqnarray}
&&
c_{\alpha}=
{\mbox e}^{(-1)^{\alpha+1}\ae_0}Z_{\alpha}(\ae_{\alpha})
{Z({\mbox{\boldmath $\ae$}})}^{-1}.
\end{eqnarray}
From (2.23),(2.24) it follows
\begin{eqnarray}
&&
m_{\alpha}=c_{\alpha}{\cal F}_{\alpha}^{(1)}(\ae_{\alpha}).
\end{eqnarray}

The equation (2.22) for unknown difference of chemical potential can be
solved analytically
\begin{eqnarray}
&&
\Gamma_0=-{\cal I}m_0+\frac{1}{2}\ln \frac{1+m_0}{1-m_0}
\frac{Z_2(\ae_2)}{Z_1(\ae_1)}.
\end{eqnarray}
Using (2.9) one can obtain following result for the free energy:
\begin{eqnarray}
\frac{1}{N} F ({\bf c, h}) =
\sum_{\alpha=1,2}c_{\alpha}\vartheta_{\alpha} +\ln 2
+\frac{1}{8}(V_{11}+V_{22}+2V_{12})
\nonumber\\
+\frac{1}{4}(V_{22}-V_{11})m_0
+\frac{1}{2}{\cal I}m_0^2
-\frac{1}{2}\ln (1-m_0^2)
-\frac{m_0}{2}\ln\frac{1+m_0}{1-m_0}
\nonumber\\
-\frac{1}{2}\sum_{\alpha,\beta=1,2}
{\cal I}_{\alpha,\beta}m_{\alpha}m_{\beta}
-\frac{m_0}{2}\ln\frac{Z_2(\ae_2)}{Z_1(\ae_1)}
+\frac{1}{2}\ln Z_1(\ae_1)Z_2(\ae_2).
\end{eqnarray}
The system (2.12) in a case ${\cal I}_{11}={\cal I}_{22}$ and
$\Gamma_1=\Gamma_2$ can be solved analytically with the following result:
\begin{eqnarray}
&&
T_b=\frac{-2{\cal I}m_0}
{\ln \frac{1-m_0}{1+m_0}\frac{Z_2({\ae_2})}{Z_1({\ae_1})}}.
\end{eqnarray}
However, $\ae_1$ and $\ae_2$ is to find from equations (2.11).
The expression for spinodal temperature can be found exactly only for
non-magnetic case:
\begin{eqnarray}
&&
T_s={\cal I}(1-m_0^2)=4{\cal I}c_1c_2.
\end{eqnarray}

The pair $CFs$ we obtain on the basis of the relation (2.8) (the details see
in \cite{ylsd}) and this procedure leads to expression for matrix of $CF$
$\hat \Omega_{\Theta}^{(2)} ({\bf q}) =
\{ \Omega^{(2)}_{\Theta,\delta_1 \delta_2}({\bf q}) \}$
%
% 16
\begin{eqnarray}
&& \hat \Omega_{\Theta}^{(2)} ({\bf q}) = [ \hat 1 - \hat f^{(2)}(q)
\hat {\cal I} (q)]^{-1} ~\hat f^{(2)} (q),
\nonumber\\
&& f^{(2)}_{\delta_1 \delta_2} ({\bf q}) = \Bigl( \frac{\delta}{\delta
\ae_{i_1 \delta_1}}\cdot \frac{\delta}{\delta \ae_{i_2 \delta_2}} f (\{ \ae
\}) {/}_{\ae_i =\ae} \Bigr) ({\bf q}),
\nonumber \\
&& \Bigl( \langle x_{i1} x_{j1} {\rangle}^c {\Bigr)}_{({\bf q})}
{=} \frac{1}{4} \Omega_{\Theta,00}^{(2)} ({\bf q}),
~\Bigl( \langle x_{i1} \,
x_{j\alpha}
s_{j\alpha} {\rangle}^c {\Bigr)}_{({\bf q})}{=}
\frac{1}{2} \Omega_{\Theta,0 \alpha}^{(2)} ({\bf q}).
\end{eqnarray}
Here $f^{(2)}_{\delta_1 \delta_2} ({\bf q})$ is the irreducible part of $CF$
$\hat \Omega_{\Theta}^{(2)} (q)$ with respect to interaction $\hat {\cal I}
({\bf q})$. It is obtained by double differentiation of irreducible
part$f({\mbox{\boldmath $\ae$}})$ with respect
to nonuniform fields $\ae_{i_1 \delta_1}, \ae_{\i_2 \delta_2}$. Within $TTA$
at $T \geq T_m$ $\hat f^{(2)}$ is independent of ${\bf q}$ and has form
%
% 17
\begin{eqnarray}
&&
f^{(2)}_{\delta_1 \delta_2}({\bf q}) =
\langle F^{(2)}_{\delta_1 \delta_2} ({\mbox{\boldmath $\ae$}}
+{\mbox{\boldmath $\sigma$}})
\rangle \stackrel{MFA}{\Longrightarrow} F^{(2)}_{\delta_1 \delta_2}
({\mbox{\boldmath $\ae$}} ),
\end{eqnarray}
%
%
% 18
\begin{eqnarray}
&&
\left. \begin{array} {ll}
F^{(2)}_{00}({\mbox{\boldmath $\ae$}})
\equiv 4c_1 c_2; F^{(2)}_{\alpha 0} = - (-1)^{\alpha} m_{\alpha}
2 c_{\alpha} \\
F_{\alpha \beta}^{(2)} ({\mbox{\boldmath $\ae$}}) = c_{\alpha}
\frac{ Z_{\alpha}^{(2)} (\ae_{\alpha}) } {Z({\mbox{\boldmath $\ae$}}) }
\delta_{\alpha,\beta}
- c_{\alpha} c_{\beta}
m_{\alpha} m_{\beta}.
\end{array}
\right \} MFA
\end{eqnarray}
\vspace{2mm}

From matrix expression for $\hat \Omega_{\Theta}^{(2)} (q)$ (2.30)
it can be found
%
% 19, 20
\begin{eqnarray}
&&
\Omega_{\Theta,00}^{(2)} ({\bf q}) =
[1 - \Phi_{00} ({\bf q}) V({\bf q}) ]^{-1}
\Phi_{00} ({\bf q}),
\nonumber \\
&&
\Omega_{\Theta}^{(2)ss} ({\bf q}) =
[1 - \hat \Phi^s ({\bf q}) {\cal I}^s({\bf q}) ]^{-1}
\hat \Phi^s ({\bf q}),
\end{eqnarray}
\begin{eqnarray}
\Phi_{00} ({\bf q}) {=} f^{(2)}_{00}({\bf q}) {+} \sum_{\alpha_1\alpha_2}
 f^{(2)}_{0\alpha_1}({\bf q})
\Bigl( \hat {\cal I}^s ({\bf q}) [1{-} \hat f^s ({\bf q})
\hat {\cal I}^s ({\bf q})]^{-1} \Bigl)_{\alpha_1 \alpha_2}
f^{(2)}_{\alpha_2 0}
({\bf q}),
\nonumber \\
\Phi^s_{\alpha \alpha '} ({\bf q}) = f^{(2)}_{\alpha \alpha '} ({\bf q})
+ f^{(2)}_{\alpha 0} ({\bf q}) V ({\bf q}) [1- f_{00} ({\bf q})
 ({\bf q})]^{-1} f_{0 \alpha '}^{(2)} ({\bf q}),
\nonumber
\end{eqnarray}
\vspace{-3mm}
\begin{eqnarray}
&&
\hat f^s ({\bf q}) = \{ f_{\alpha\beta}^{(2)} ({\bf q}) \},
\quad {\cal I}^s ({\bf q}) = \{ {\cal I}_{\alpha\beta} ({\bf q})
\}.
\end{eqnarray}
At the temperature of second order phase transition the  $CFs$ of system
diverge at certain values ${\bf q}^{\ast}$ (we consider here only case
${\bf q}^{\ast} = 0$).

In the case $T > T_m$ the $f_{0\alpha}^{(2)} ({\bf q}) \equiv 0$ and
$\Phi_{00}({\bf q}) = f_{00}^{(2)} ({\bf q})$,
$\Phi_{\alpha\alpha '}^{s} ({\bf q}) = f_{\alpha\alpha '}^{(2)}({\bf q}) $.
Within $MFA$ from first expression we obtain the $T_s$ (the influence of
magnetic subsystem is absent) and from the second one the equation for $T_m$
($T_m$ is the temperature of magnetic transition)
%
% 21, 22
\begin{eqnarray}
&& T_s = 4 c_1 c_2 {\cal I} (0), \nonumber\\
&& 2 T_m = c_1 T_{11} + CT_{22} - \{ (c_1 T_{11} - c_1 T_{22})^2 +
4c_1 c_2 ~T^2_{12} \}^{1/2}, \\
&& T_{\alpha \beta} = {\cal I}_{\alpha \beta}
\sqrt {{\cal F}_{\alpha}^{(2)}~
{\cal F}_{\beta}^{(2)} } .
\end{eqnarray}
%

%\section*{4. Results of numerical investigation}
\section{Results of numerical investigation}

The Hamiltonian (2.1) corresponds to the most generalized model of
$M$-com\-ponent magnet.  There is no possibility at present to obtain any
analytical solutions in this case. Therefore, in this section we represent
some numerical results performed for particular cases of one-component Ising
model, binary alloys and lattice gas. Here the numerical calculations become
simpler.

%\subsection*{4.1. Ising model}
\subsection{Ising model}
\setcounter{equation}{0}

The case of Ising model will correspond to $M=1$. The ionic part of the
Hamiltonian becomes constant and does not influence on the thermodynamical
and correlation functions of the system:
\begin{eqnarray}
&&
{\cal H}=\sum_{i=1}^N\Gamma_1S_{i1}+
\frac{1}{2}\sum_{i,j=1}^N{\cal I}_{11}S_{i1}S_{j1}+const
\end{eqnarray}

For calculation of free energy, magnetization, two-tail we can use formulae
(2.13)-(2.20) and put $\alpha,\beta=1$. In this case we have to solve system
of only two equations for order parameter $m_1$ and fluctuating parameter
$\lambda_{11}^{(2)}$. Therefore, one can write the following expression for
generating function in the framework of two-tail approximation (2.13)-(2.18)
that corresponds to free energy in the case of Ising model:

%
%\renewcommand{\theequation}{4.\arabic{equation}}
%
% 1
\begin{eqnarray}
&&- \frac{\beta}{N} F (\Gamma) =
- \frac{1}{2} {\cal I}_{11} m^2_{1}
+ \langle F^{(0)} (\ae_1 +
\sigma_1) \rangle
-\frac{1}{2}
\lambda^{(2)}_{11} \langle F^{(2)}_{11} (\ae_1 +
{\sigma_1}) \rangle
\nonumber\\
&&- \frac{1}{2N} \sum_{q} \ln [1 -
{\cal I}_{11} (q) \langle F^{(2)}_{11} (\ae_1 +
\sigma_1) \rangle ].
%\nonumber
\end{eqnarray}
The equations for order parameters $m_1$ and two-tail $\lambda_{11}^{(2)}$
follow from the equations (2.17)-(2.18) :
%
% 2, 3
\begin{eqnarray}
&& m_1 =  \langle {F}^{(1)}_1 \{\ae_1 + \sigma_1\}
{\rangle},\\
&& \lambda_{11}^{(2)} = \frac{1}{2N} \sum_{q} \frac{{\cal I}_{11}(q)}
{1 - {\cal I}_{11}(q) \langle F^{(2)}_{11} \{\ae_1 + \sigma_1\}
{\rangle}}.
\end{eqnarray}
Here the averaging is performed with distribution function :
%
% 4
\begin{eqnarray}
&&
\rho_2 \{ \sigma\} = \int \frac{d \zeta}{2\pi} e^{i\zeta \sigma}
e^{-\frac{1}{2}\zeta^2 \lambda_{11}^{(2)}} =\frac{1}{ \sqrt{2\pi
\lambda_{11}^{(2)} }} e^{-\frac{\sigma^2}{ 2
\lambda_{11}^{(2)}}}.
\end{eqnarray}
%
%In the works of Onyszkiewicz at all \cite{onysz} Gaussian field
%approximation (GFA) was suggested. It
%can be obtained expanding
%$\ln [1-{\cal I}(q)
%\langle {\cal F}^{(2)} \{\ae + \sigma\} {\rangle}_{\rho_2}]$ in (3.1)
%and $1/ (1- {\cal I}(q) \langle {\cal F}^{(2)} \{\ae + \sigma\}
%{\rangle}_{\rho_2}$in (3.3) in ${\cal I}(q)$ up to the terms of second
%order. In the result :
GFA gives the following results for the free energy and equation for
two-tail $\lambda_{11}^{(2)}$ (see (2.19), (2.20)):
%
% 5, 6, 7
\begin{eqnarray}
&&- \frac{\beta}{N} F (\Gamma) =
- \frac{1}{2} {\cal I}_{11} m^2_{1}
+ \langle F^{(0)} (\ae_1 +
\sigma_1) \rangle
-\frac{1}{2}
\lambda_{11}^{(2)} \langle F^{(2)}_{11} (\ae_1 +
{\sigma_1}) \rangle
\nonumber\\
&&+\frac{1}{4}\langle F^{(2)}_{11} (\ae_1 +{\sigma_1}) \rangle
\left[
\langle F^{(2)}_{11} (\ae_1 +{\sigma_1}) \rangle
\frac{1}{N}\sum_q{\cal I}^2_{11}(q)-2\lambda_{11}^{(2)}
\right],
\end{eqnarray}

\begin{eqnarray}
&& m_1 = \langle F^{(1)} \{\ae_1 + \sigma_1\}
{\rangle}_{\rho_2}, \\
&& \lambda_{11}^{(2)} = \langle F^{(2)}_{11} \{\ae_1 +
\sigma_1\} {\rangle}_{\rho_2}\frac{1}{N} \sum_{q} {\cal I}^2_{11} (q).
\end{eqnarray}
Neglecting the fluctuation term $-\frac{1}{2} \sum\limits_{ij}{\cal I}_{ij}
\Delta S_i \Delta S_j$ in the Hamiltonian we get the MFA approximation. This
concerns the case $\lambda_{11}^{(2)} =0$,
%\linebreak
$-\beta F(\Gamma)/N$ $= -\frac{1}{2}
{\cal I}_{11}
m_1^2 + F^{(0)}\{\ae_1\}$ and $m_1
= F^{(1)}\{\ae_1\}$.

The numerical calculations were performed for the model with nearest
neighbours interaction :
%
% 8
\begin{equation}
{\cal I}_{11} (q) = I\sum_{\alpha=1}^{3} \cos q\alpha.
\end{equation}

The results of this investigation can be seen in Fig.1, 2. In Fig.1 the
free energy of the system as a function of order parameter for various
temperatures is depicted. It is known, that we get first order
phase transition in the framework of two-tail approximation. This can be
seen in Fig.1. The curve for $T/{\cal I}(0)=0.72$ has two minima with equal
energy. Although the GFA is more rough than two-tail approximation, it gives
us a second order phase transition at $T/{\cal I}(0)=0.86$. The results of
temperature dependences of magnetization in different approximation are
plotted in Fig.2.
\begin{figure}[hp]
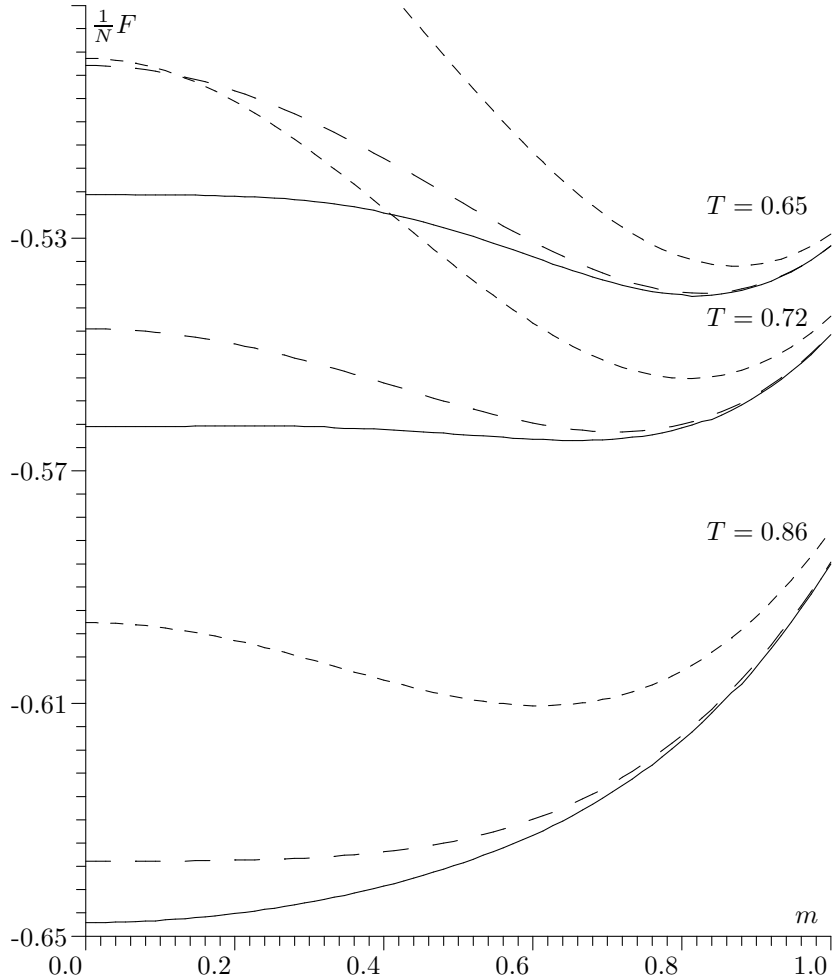

%\begin{center}

\hspace{-1cm}
\input F26E1.PIC

%\end{center}
\caption{The free energy $\frac{1}{N}F$ as a function of magnetization $m$
for various temperature (solid curves - two-tail approximation, long
dashed curves - GFA, short dashed curves - MFA)} \end{figure}
\begin{figure}[thp]
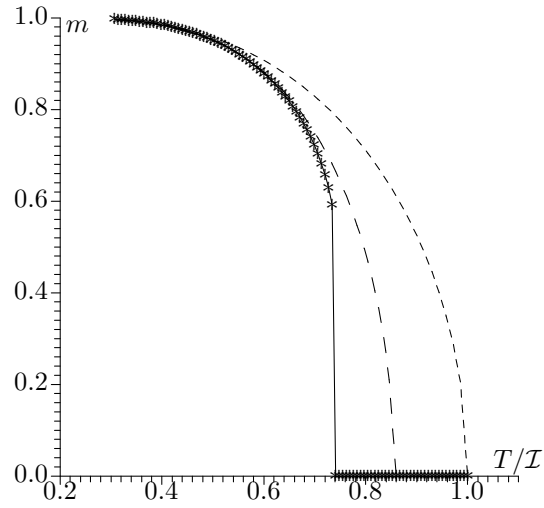

\begin{center}
\input F26E2.PIC
\end{center}
\caption{The magnetization $m$ as a function of temperature $T$  (solid
curve - two-tail approximation, long dashed curve - GFA, short dashed
curve - MFA)}
\end{figure}
\begin{figure}[ph]
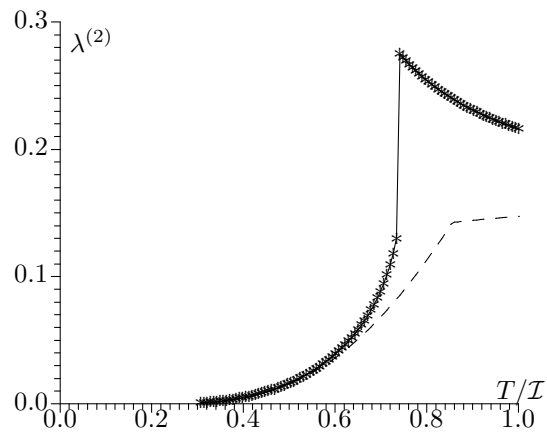

\begin{center}
\input F26E3.PIC
\end{center}
\caption{The two tail $\lambda$ as a function of temperature $T$  (solid
curve - two-tail approximation, dashed curve - GFA)}
\end{figure}

The obtained results show that the two-tail approximation predicts a
non-physical beha\-vi\-our in the close vicinity of the critical
temperature. This may lead to the non-controlled errors by investigation of
binary alloys and lattice gas. Therefore, for these systems we shall
restrict only GFA and compare results with MFA.

%\subsection*{4.2. Binary alloy}
\subsection{Binary alloy}

The Hamiltonian of the non-magnetic binary alloy follows from (2.1) if we put
${\cal I}_{11} (i-j) = {\cal I}_{12} (i-j) ={\cal I}_{22} (i-j) = 0,
\Gamma_{i1} = \Gamma_{i2} = 0$:
\begin{eqnarray}
&&
{\cal H}=  {\cal H}_x = \sum_{\alpha=1}^2   \sum_{i = 1}^N
\varepsilon_{\alpha} x_{i \alpha} + \frac{1}{2} \sum_{\alpha,\beta = 1}^2
\sum_{i,j=1}^N V_{\alpha\beta}(i-j) x_{i \alpha} x_{j\beta}.
\end{eqnarray}
The MFA and GFA may be easily reformulated for this case using formulae of
previous section.
Our numerical calculations concern the system with nearest  neighbours
interaction. As may be proved in section 2 the critical temperatures of
binodal and
spinodal decay depends only on ${\cal I} = V_{11} + V_{22} - 2V_{12}$ and
are symetrical relatively substitution $c_2$ instead of $c_1$. This can be
seen in Fig. 4, where binodal and spinodal temperatures as a function of
$c_1$ are depicted. The comparison of GFA and MFA approximation results
show that  GFA approximation is essentially better that MFA in the
vicinity of $c_1 =0.5$. When $c_1 \to 0$ or $c_1 \to 1$, the difference
disappears.  It should be noticed that two-tail approximation  will give
non-physical results namely in the vicinity  of $c_1= 0.5$.
\begin{figure}
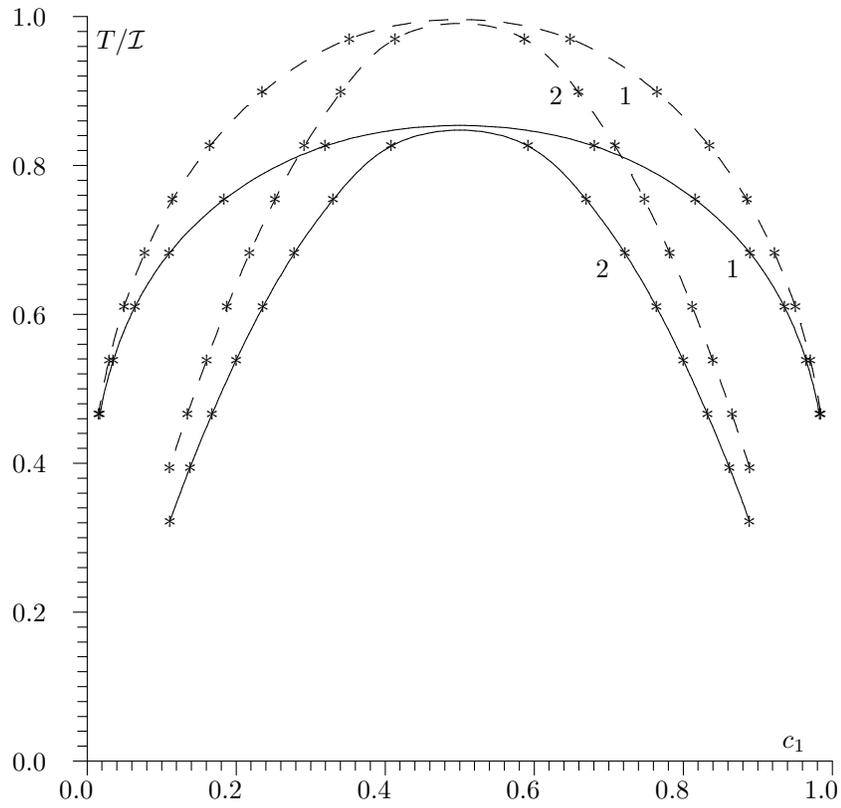

\begin{center}
\input F26E4.PIC
\end{center}
\caption{
Phase diagrams for the model (3.9): 1- correspond to the binodal
temperature, 2- spinodal temperature (solid line - GFA, dashed line -
MFA)}
\end{figure}

%\subsection*{4.3. Lattice gas}
\subsection{Lattice gas}

We shall get the Hamiltonian of the non-magnetic lattice gas, if we put
$V_{12}(i-j) = V_{22} (i-j) = 0$ in case of binary alloy:
%
%13
\begin{eqnarray}
&&
{\cal H}=\sum_{i=1}^N \mu x_i+\frac{1}{2}\sum_{i,j=1}^N V(i-j)x_ix_j.
\end{eqnarray}
Indeed, the Hamiltonian (3.11) may describe the gas. Let us consider the
system in volume ${\cal V}$ and divide volume on $N={\cal V}$ equal cubic
cells. We will use the condition that more than one particle can
not be situated in one site. Therefore, we get a lough model of a gas with
repulsion between particles on short distance and model interaction for long
distance.

Here we neglect the kinetic energy of the particles. But being taken
into account it leads only to the renormalization of the chemical potential
%
%14
\begin{eqnarray}
&&
\mu \rightarrow \mu+\frac{3}{2}\ln
\left(2\pi h^{-2}mkT\right),
\end{eqnarray}
where m - particle's mass.

Similarly to (2.5), the Hamiltonian (3.13) can be transformed to
\begin{eqnarray}
&&
{\cal H}=E(\mu)+\sum_{i=1}^N\Gamma_0S_{i0}+
\frac{1}{2}\sum_{i,j=1}^N{\cal I}(i-j)S_{i0}S_{j0},
\end{eqnarray}
where $E(\mu)=\frac{1}{2}\mu+\frac{1}{8}V$,
$\Gamma_0=\frac{1}{2}\mu+\frac{1}{4}V$, $4{\cal I}(i-j)=V(i-j)$.\\
It is known from thermodynamics, that thermodynamical potential of the
system $\Omega(\mu,T)=-p{\cal V}$, where $p$ is pressure and ${\cal V}$
is volume of the system.
The average of $x_i$ that equal $n=1/v$, plays the role of gas density.
The state equation can be
written in the following form
\begin{eqnarray}
&&
p{\cal V}=\Omega(\mu(n),T) \;
{\mbox or} \;
p=\frac{1}{\cal V}\Omega(\mu(n),T).
\end{eqnarray}

In MFA the state equation can be obtained in the analytical form. The
expression for TP follows from (2.21) and has the following form:
\begin{eqnarray}
&&
\frac{1}{N}\Omega(\mu,T)=
\frac{1}{2}\mu+\frac{1}{2}V-\frac{1}{2}{\cal I}m_0^2
+\ln 2\cosh (\Gamma_0+{\cal I}m_0),
\end{eqnarray}
where $\Gamma_0(m_0)=-\beta{\cal I}m_0+
\frac{1}{2}\ln \frac{1+m_0}{1-m_0}$,
$\mu=2\Gamma_0-\frac{1}{2}{\cal I}$.
If we take into account that $m_0=2n-1$, we can obtain state equation in
variables $p$, $n$, $T$:
\begin{eqnarray}
&&
\frac{p}{T}=-\beta an^2+\Phi(n),
\end{eqnarray}
where $\Phi(n)=\ln \frac{1}{1-n}$.
It should be noted that for van der Waals state equation
$\Phi(n)=\frac{n}{1-bn}$.

We have also obtained isoterms in GFA.
Since the Hamiltonian of the lattice gas is identical to the
Hamiltonian of the binary alloy with $V_{12}(i-j)=V_{22}(i-j)=0$, we can use
all approximation formulated in the previous subsection to its
investigation.

In the Fig.5 the isoterms for the lattice gas with nearest neighbours
\begin{figure}
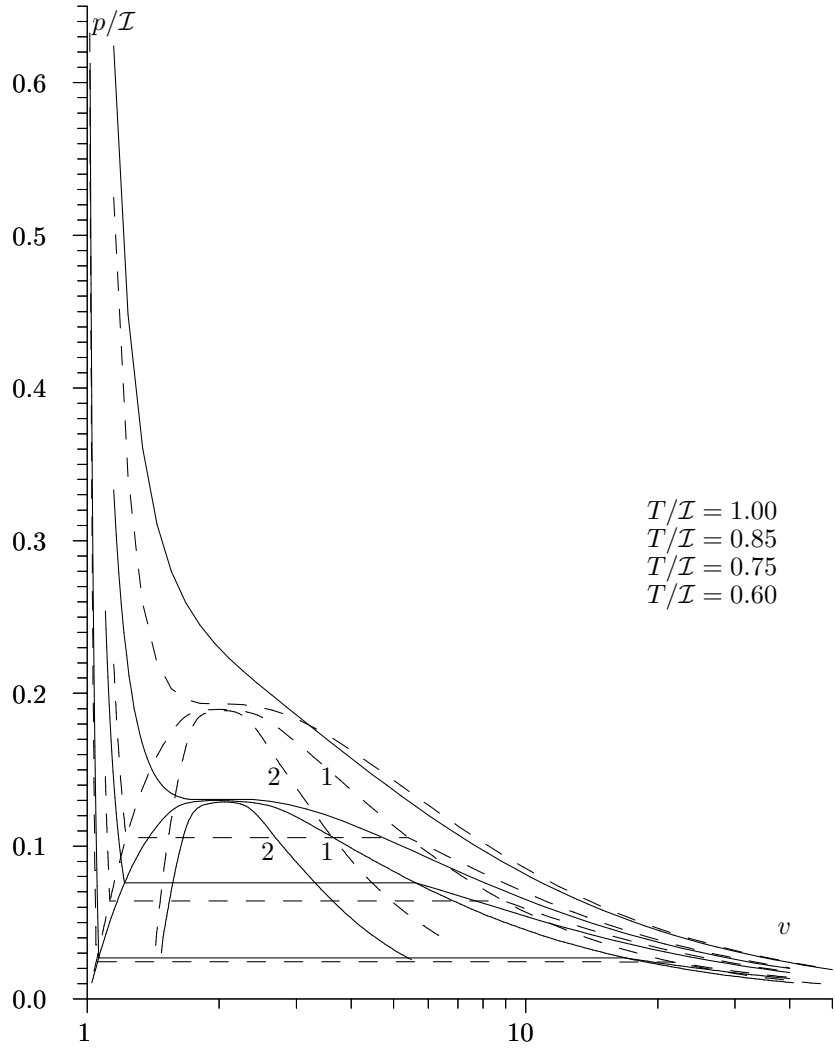

\begin{center}
\input F26E5.PIC
\end{center}
\caption{The
$p-V$ diagram for the lattice gas by various temperature
(solid
curves - two-tail GFA, dashed curves - MFA; 1 - binodal, 2
- spinodal)}
\end{figure}
interaction are depicted. The dashed curve corresponds to the MFA
and solid curve for the GFA. It can be seen that
values of the pressure $p$ and binodal and spinodal temperature of the
system obtained within GFA are smaller than those obtained within MFA.

\section{Conclusions}
In this paper we have considered the $M$-component Ising model with site
disorder.
Several approximation for this model have been formulated here. Namely, we
have obtained the expressions for the thermodynamical potential, free
energy and pair correlation functions within TTA, GFA and MFA.

The numerical calculations were performed for some simpler models:
one-component Ising model, non-magnetic binary alloy and lattice gas.  For
example, we have chosen the nearest neighbours interaction.
For the one-compo\-nent Ising model we calculated the free energy as a
function of the magnetization, temperature dependences of the magnetization
$m(T)$ and the two-tail $\lambda^{(2)}$ (fluctuating parameter) within TTA,
GFA, MFA. For non-magnetic binary alloy we obtained phase diagrams (binodal
and spinodal curves) within GFA and MFA. For lattice gas the isoterms and
coexistence curves are depicted.\\

This work was supported in part by the International Soros Science Education
Program (ISSEP) through grant  No. PSU062015.

%\clearpage

\end{document}